\documentclass[a4paper,11pt]{article}
\pdfoutput=1 

\usepackage{jheppub} 

\usepackage[T1]{fontenc} 
\usepackage{xcolor}

\title{\boldmath Thick Branes in Extra Dimensions and Suppressed Dark Couplings}

\author[a]{Ricardo G. Landim,}
\author[a]{Thomas G. Rizzo}

\affiliation[a]{SLAC National Accelerator Laboratory, 2575 Sand Hill Rd., Menlo Park, CA 94025 USA}

\emailAdd{rlandim@slac.stanford.edu}
\emailAdd{rizzo@slac.stanford.edu}

\abstract{The nature of dark matter (DM) and how it may interact with the various fields of the Standard Model (SM) remains a mystery. 
In this paper we show that the interaction between new light dark matter mediators and the SM particles can be naturally suppressed if one employs a single, flat extra 
dimension (ED).  In this setup, the SM fields are localized in a finite width `fat' brane, similar to models of Universal Extra Dimensions (UED), while DM, in turn, is confined to a 
thin brane at the opposite end of the ED interval. Including brane localized kinetic terms on the fat brane for the mediator fields, the resulting coupling between the SM and these light 
mediators can be several orders of magnitude smaller than the corresponding ones between the mediators and DM which we assume to be a typical gauge coupling. We investigate 
the implications of this scenario for both vector (\textit{i.e}, dark photon, DP) 
and scalar mediator fields in the 5-D bulk. In this setup kinetic mixing, which is usually employed to suppress light mediator couplings, is not required. Here we assume that the SM 
particles couple to the DP via their $B-L$ charges while the DP couples to the DM via a dark charge. Both the vector DP couplings and the corresponding Higgs portal couplings with 
the SM are shown to be natural small in magnitude with a size dependent on ratio of the 5-D compactification radius, $R^{-1} \sim 0.1-1$ GeV, and the SM brane thickness, 
$L^{-1} \sim 2-10$ TeV, a range chosen to avoid LHC and other experimental constraints. In this framework one can obtain the observed value of the DM relic abundance for a wide range 
of parameter choices, while the constrains due to direct DM detection and the invisible width of the Higgs do not impose significant challenges to the model. Finally, this mechanism can 
lead to distinct signatures in both present and upcoming experiments as it combines some common features of UED and DP models in a single ED setup.}

\begin{document} 
\maketitle
\flushbottom

\section{\label{sec:intro}Introduction}

The nature of the dark matter (DM) remains a puzzling challenge to modern cosmology and particle physics. Although Weakly Interacting Massive Particles (WIMPs) are the most well-known  
candidates (see Ref.\cite{Arcadi:2017kky} for a recent review), the lack of any positive signatures lead us to consider different scenarios, both from the theoretical and experimental 
point-of-view. Among the many possibilities, the interaction of DM with Standard Model (SM) particles through a new mediator field is a promising avenue of approach to the DM problem  
considering the diverse set of existing and planned direct detection, indirect detection and accelerator experiments \cite{Battaglieri:2017aum}.  A widely studied example of the approach is 
the dark photon model (DP) \cite{Feldman:2006wd,Feldman:2007wj,Pospelov:2007mp,Pospelov:2008zw,Davoudiasl:2012qa,Davoudiasl:2012ag,Essig:2013lka,Izaguirre:2015yja} (see Ref.  \cite{Curtin:2014cca} for a review) wherein a new, relatively light, dark $U(1)_D$ gauge field interacts with us via kinetic mixing with the SM 
hypercharge $U(1)_Y$ field \cite{Holdom:1985ag,Holdom:1986eq,Dienes:1996zr,DelAguila:1993px,Babu:1996vt,Rizzo:1998ut}.

Many extensions of the SM appear by employing extra dimensions (ED). As is well-known, the entire SM can be embedded in ED and this is most explicitly the case in the Universal Extra 
Dimension (UED) models, with either one \cite{Appelquist:2000nn} or two ED \cite{Burdman:2005sr,Ponton:2005kx}, for example. In UED models the whole SM content is thus 
promoted to fields which propagate in the compact ED thus having Kaluza-Klein (KK) excitations. In 4-D, the lightest/zero mode of each KK tower of states is then identified with the 
correspondent SM particle, and the KK tower of excitations are heavier partner particles with the same spin as the zero mode. Searches for Supersymmetric (SUSY) \cite{Aad:2015mia,TheATLAScollaboration:2013uha} particles 
at the LHC can then be re-interpretted to constrain the UED compactification radius, $L$, since both classes of models can have qualitatively similar phenomenology. Those searches 
imposed the current lower bound on the UED radius $L^{-1} >1.4-1.5$ TeV  (for $\Lambda L\sim 5-35$, where $\Lambda $ UED is the cutoff scale)  
\cite{Deutschmann:2017bth,Beuria:2017jez,Tanabashi:2018oca}. Other ED models have been employed to address multiple issues in particle physics such as the gauge hierarchy 
\cite{Antoniadis:1990ew,Dienes:1998vh,Antoniadis:1998ig,ArkaniHamed:1998rs,Randall:1999ee,Arkani-Hamed:2016rle} and flavor problems 
\cite{Agashe:2004cp,Huber:2003tu,Fitzpatrick:2007sa}. In the context of ED, the DP model has been recently embedded in a flat, single ED, together with a DM candidate being either 
a scalar or a fermion \cite{Rizzo:2018ntg,Rizzo:2018joy}. In this scenario, the gauge boson in 4-D acquires a mass due to the breaking of $U(1)_D$ by boundary conditions (BC), thus no dark 
Higgs field is explicitly needed to generate the DP mass.

For the usual scalar singlet DM candidate, symmetric under $\mathbb{Z}_2$ \cite{Silveira:1985rk,McDonald:1993ex,Burgess:2000yq}, the interaction with the SM is only through the 
Higgs portal, whose implications have been examined in multiple contexts \cite{Bento:2000ah,Bertolami:2007wb,Bento:2001yk,MarchRussell:2008yu,Biswas:2011td,Costa:2014qga,Eichhorn:2014qka,Khan:2014kba,Queiroz:2014yna,Kouvaris:2014uoa,Bhattacharya:2016qsg,Bertolami:2016ywc,Campbell:2016zbp,Heikinheimo:2016yds,Kainulainen:2016vzv,Nurmi:2015ema,Tenkanen:2016twd,Casas:2017jjg,Cosme:2017cxk,Heikinheimo:2017ofk,Landim:2017kyz}. The size of the Higgs portal coupling is highly constrained by 
the  bound on the invisible Higgs width for a general scalar, while for a DM candidate obtaining the observed relic density while also satisfying direct detection searches forces the parameter 
space to be even more highly constrained \cite{Duerr:2015aka,Athron:2017kgt}, requiring a highly tuned and very small Higgs portal coupling. The simplest version of this scenario 
is now almost, if 
not completely, excluded by these multiple constraints from several observations and experiments \cite{Djouadi:2011aa,Cheung:2012xb,Djouadi:2012zc,Cline:2013gha,Endo:2014cca,Goudelis:2009zz,Urbano:2014hda,Akerib:2015rjg,He:2016mls,Escudero:2016gzx,Ade:2015xua,Cline:2013fm,Slatyer:2015jla,Ackermann:2015zua,Akerib:2016vxi,Tan:2016zwf,Agnese:2014aze,Aprile:2012nq,Aartsen:2012kia,Aartsen:2016exj}.

The small values of both the Higgs portal and DP couplings with the SM can appear naturally through an ED effect, as we shall see below. Here we will assume that the new $U(1)_D$ 
gauge symmetry produces an interaction between SM and DM since the SM particles carry a  $B-L$ charge while the DM has a dark charge. As we will see, for the DP in the bulk, there 
is no need for kinetic mixing to suppress this interaction. To be specific, we will introduce  a single, flat ED, a DM candidate confined to one thin brane at one interval boundary while the 
SM is contained in a brane 
of finite thickness at the opposite end of the interval. The effect of this `fat' brane is to suppress the couplings with the SM, leading to a natural small gauge or Higgs portal coupling. 

This paper is organized as follows. Sect. \ref{General Framework Setup} describes the basic framework that will be employed in this setup. In Sect. \ref{sec:gauge} we consider a gauge 
field in the bulk 
and analyze the resulting couplings with the SM and DM. Also, we examine the constraints on the SM interactions with the DM particle from both direct and indirect observations. In 
Sect. \ref{sec:scalar} we turn our attention to the case of a scalar mediator field in the bulk, derive the appropriate expressions to obtain the Higgs portal coupling and, in turn, analyze 
the invisible Higgs decay constraints in the ED context. Finally, Sect. \ref{sec:discussion} is reserved for conclusions.

\section{Framework}\label{General Framework Setup}

We will consider the case of a single flat ED, represented by an interval, with one thin brane localized at the $y=0$ boundary where the DM candidate is confined, and a `fat' (\textit{i.e.}, thick) 
brane lying between $y=\pi r$ and $y=\pi R$, with a width $\pi(R-r)\equiv \pi L$, at the opposite boundary. The SM is contained within this fat brane and it is assumed that  $L\ll R$. 
This scenario in which the SM experiences a TeV-scale ED is similar to models of UED \cite{Appelquist:2000nn},  in which the SM content is promoted to fields which propagate through this compact ED of size $\pi L$, thus having KK excitation, whose eigenfunctions are $\sim \cos n y /L $ (or $\sim \sin n y/L$), where $n$ labels the KK tower state.\footnote{SM has also been assumed to be localized in branes in different contexts, as for instance in Refs. \cite{ArkaniHamed:1998rs,Dvali:1998pa}, while in Refs.  \cite{DeRujula:2000he,Georgi:2000wb} gravitons or fermion are localized in fat branes.} 

Throughout the paper,  where needed to exemplify our calculations, we will consider four specific benchmark models (BM), whose assumed set of values for the compactification radius $R$ and the size of the UED $L$ are presented in Table \ref{table:param}. We 
note that these are merely representative and that many other choices of these parameters would yield results qualitatively similar to at least one of these cases. 
 
\begin{table}[h] 
       \centering 
           \begin{tabular}{c| c c c c}
           \hline\hline
    BM  &   I &  II &  III &  IV\\
   \hline
    $R^{-1} $  & $1 $ GeV &  $1$ GeV  & $100$  MeV & $100$ MeV  \\
    $L^{-1}$  &$2$ TeV  & $10$ TeV  & $2$ TeV & $10$ TeV  \\
         \hline\hline \end{tabular}
 \caption{The different sets of parameters used in BM for the present study.}   \label{table:param}
\end{table}

\section{Bulk Gauge Field Mediator}\label{sec:gauge}
We will first consider an abelian gauge field $V_A, ~A=0-3,5$ in the bulk, which interacts with both the DM and the SM. In the absence of kinetic mixing, the DP couples with DM and SM 
through the usual covariant derivative, however, it should not couple to the SM Higgs since such a coupling would influence, \textit{e.g},  the $Z$ boson mass and other SM precision 
electroweak predictions which in turn are very well constrained \cite{Tanabashi:2018oca}. One consistent way to introduce this coupling with SM particles is to use an anomaly-free 
symmetry under which 
baryons and/or leptons are charged, while the SM Higgs and gauge bosons are not. As is well-known, four symmetries may play this role without the presence of additional SM fermion 
fields (beyond RH--neutrinos) to cancel anomalies: the difference between baryon and lepton numbers ($U_{B-L}$) and the three differences between the lepton numbers 
($U_{L_\mu-L_e}$, $U_{L_e-L_\tau}$ and $U_{L_\mu-L_\tau}$) \cite{Foot:1990mn,He:1990pn,He:1991qd,Bauer:2018onh}. DM, on the other hand, is assumed to not have any SM charges, 
in particular, a non-zero $B-L$ charge \footnote{Although it might have, as in the models of asymmetric dark matter \cite{Petraki:2013wwa}}. However, the DM does carry (a vector-like) 
dark charge $Q_D$, therefore, we will assume that the DP couples to the combination $B-L+\lambda Q_D$ where $\lambda=1$ can be chosen without loss of generality. Thus 
the covariant derivative contains  a term proportional to $\sim g_{5D}(B-L+Q_D)$, where $g_{5D}$ is 5-D the dark gauge coupling. For the SM particles $B-L\neq 0$ and $Q_D=0$, while for 
DM, $B-L=0$ by assumption and we will assume that $Q_D=1$.  No dark Higgs field is introduced as the symmetry is broken by the choice of BC. 

For completeness and to help with the numerics of the model construction we will also consider brane-localized kinetic terms (BLKT) at $y=0$, and within the thick SM 
brane \cite{Dvali:2000rx}. The action for the DP field in the presence of these BLKT terms is then given by\footnote{The induced kinetic term on the brane is effectively 4-D at distances shorter than $R$, even in the case of a thick brane, as explained in \cite{Dvali:2000rx}.} 
\begin{equation}\label{eq:actionV}
    S=\int d^4x\int_0^{\pi R} dy \left[ -\frac{1}{4}V_{AB}V^{AB}-\frac{1}{4}V_{\mu\nu}V^{\mu\nu}\cdot \delta_AR\, \delta(y)-\frac{1}{4}V_{\mu\nu}V^{\mu\nu}\cdot \delta_B R\, \theta(y)\right]\,,
\end{equation}
 where $A$ is the 5-D index  and $\theta(y)$ is a step-function for the BLKT spread inside the thick brane \cite{Dvali:2000rx}, \textit{i.e.}, 
  \begin{equation}\label{eq:theta}
      \theta(y)=\alpha \quad \text{for } \pi r < y \leq \pi R\,, \qquad \theta(y)=0\quad \text{for } y<\pi r\,,
  \end{equation}
where $\alpha>0$. Although the parameter $\alpha$  is a new energy scale in the model, it is always multiplying the parameter $\delta_B$ and, as we shall see, variations of $\delta_B\alpha$ does not have a major effect on the results. In this sense, the energy scales associated with $R$ and $L$ are much more relevant than $\alpha$.

As usual, we expand the 5-D gauge field into a KK tower of states
  \begin{equation}
      V^{\mu[5]}(x,y)=\sum_n v^{[5]}_n(y)V^{\mu[5]}_n(x)\,,
  \end{equation}
which yields the following equation of motion for $v_n(y)$ employing for convenience the $V^5=0$ gauge:
\begin{equation}\label{eq:eom}
  \partial_y^2v_n+(m_n^V)^2v_n+\delta_AR\, \delta(y)(m_n^V)^2v_n+\delta_BR\, \theta(y)(m_n^V)^2v_n=0\,.
\end{equation}
For $y<\pi r$, the step-fuction $\theta(y)=0$, leading to a simpler equation of motion  \cite{Carena:2002me}, whose solution is found after applying the correspondent BC at $y=0$ employing the familiar solution for the  equation of motion with a delta-function source (\textit{i.e.}, the continuity of the function and the discontinuity of its 
derivative)  \cite{Rizzo:2018ntg}
\begin{equation}\label{eq:v1}
    v_{1,n}(y)=N_n^V\left[ \cos (m_n^Vy)-\frac{\delta_A x_n^V}{2} \sin(m_n^V y)\right]\, \qquad 0\leq y \leq \pi r\,,
\end{equation}
where the resulting 4-D DP KK masses are given by $m_n^V=x_n^V/R$ and $N^V_n$ is a normalization factor. 
On the other hand, for $y>\pi r$, $\theta(y)=\alpha$ and we can define an effective mass parameter given by
\begin{equation}\label{massbar}
    \bar{m}_n^V=m_n^V\sqrt{1+ \delta_B\alpha R }\,,
\end{equation}
such that the wave function in this region therefore must have the following form
\begin{equation}
   v_{2,n}(y)=A_n\cos ( \bar{m}_n^V y)+B_n \sin( \bar{m}_n^V y)\,.
\end{equation}
The coefficients $A_n$ and $B_n$ are then found through the standard requirement that the wavefunction and its derivative should be continuous at the $y=\pi r$ `boundary'. After applying 
this condition we obtain (here a $'$ denotes a derivative with respect to the $y$ co-ordinate) 
\begin{equation}\label{eq:v2}
    v_{2, n}(y)=v_{1, n}(\pi r) \cos [\bar{m}_n^V(y-\pi r)]+\frac{v_{1, n}'(\pi r)}{\bar{m}_n^V} \sin[\bar{m}_n^V(y-\pi r)]\, \qquad \pi r\leq y\leq \pi R\, .
\end{equation}

\begin{figure}
    \centering
    \includegraphics[scale=0.45]{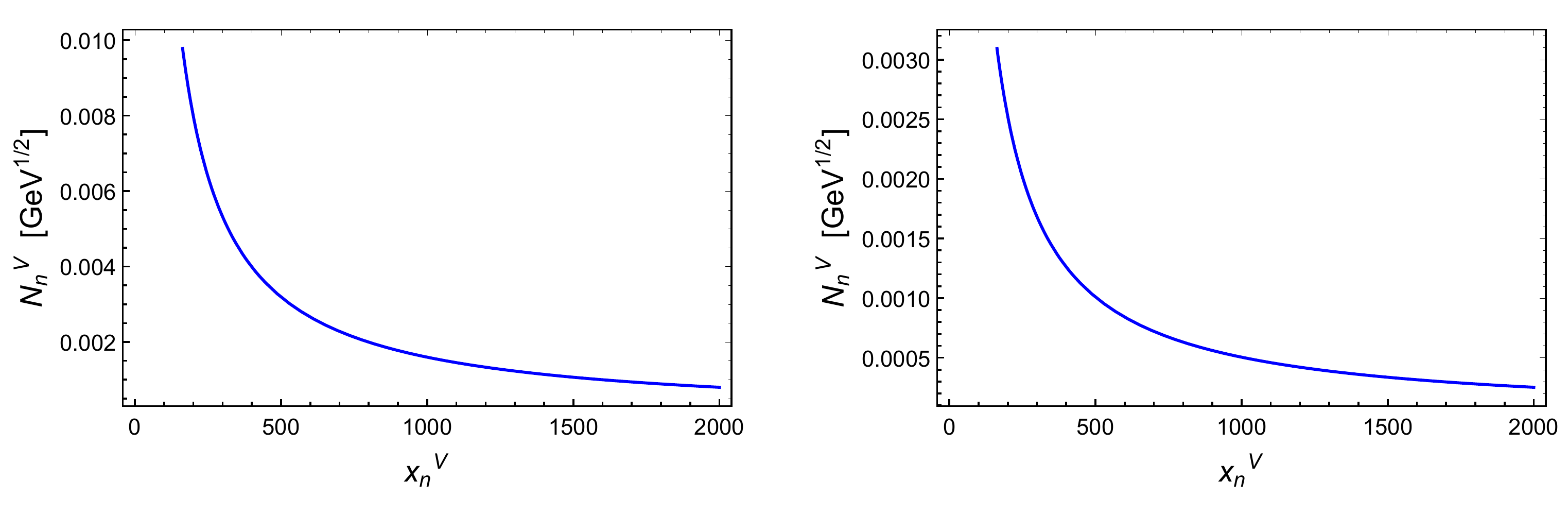}
    \caption{Normalization $N_n^V$ as a function of $x_n^V$, for BM II (left) and IV (right) with $\delta_A=\delta_B=1$ and $\alpha=1$ GeV. BM I (III) presents basically the same behavior as BM II (IV) which are not shown. }
    \label{fig:nu-plot}
\end{figure}

The presence of the BLKT inside the thick brane is responsible for the existence of this distinct wave function in this region. If $\delta_B \alpha=0$ the effective masses $ \bar{m}_n^V$ would 
reduce to the same expression as that for the physical masses $ m_n^V$ and the solution as presented in Eq. (\ref{eq:v2}) would be $v_{1,n}(y)$.
Finally, using the BLKT modified orthogonality relations
\begin{align}
    \int_0^{\pi R}dy\, [1+\delta_A R\, \delta(y)+\delta_B R \,\theta(y)]v_m(y)v_n(y)&=\delta_{m,n}\nonumber\\
     \int_0^{\pi R}dy\, \partial_y v_m(y)\partial_yv_n(y)&=m_m^V m_n^V\delta_{m,n}\,,
\end{align}
we can obtain the overall normalization factors for the wave functions:  
\begin{align}\label{eq:Nu}
\frac{2}{R}(N_n^V)^{-2}=&
\left[\frac{1}{x_n^V}- \left(\frac{\delta_A x_n^V}{2}\right)^2\right] \sin \left(\frac{2 \pi  r x_n^V}{R}\right)+  \delta_A  \cos \left(\frac{2 \pi  r x_n^V}{R}\right)+ \frac{2 \pi r}{R}\left[\left(\frac{   \delta_A  x_n^V }{2}\right)^2+1\right]
  \nonumber\\
&+\delta_A- \left[ \frac{2 \pi L}{R}+\frac{ \sin \left(\bar{m}_n^V\pi  L\right)}{\bar{m}_n^VR}\right]
\csc ^2\left(\bar{m}_n^V\pi  L\right)\left(  v_{1, n}(\pi r)\right)^2(1+ \delta_B\alpha R )  \,.
\end{align}
The behavior of Eq. (\ref{eq:Nu}) is depicted in Fig. \ref{fig:nu-plot}, for some representative BM values of $R$ and $L$ as given in Table \ref{table:param}.
The BC $v_{2, n}(\pi R)=0$ leads to the root equation which determines the physical DP KK tower masses: 
\begin{equation}\label{eq:rooteq}
  \tan(\bar{m}_n^V \pi L)=-\bar{m}_n^V\frac{v_{1, n}(\pi r)}{v_{1,n}'(\pi r)}\,,
\end{equation}
whose solutions determine the physically allowed (quantized) values of $x_n^V$. All previous results for a thin brane \cite{Rizzo:2018ntg,Rizzo:2018joy} are recovered in the limit that 
$\delta_B\rightarrow 0$ and $r\rightarrow R$. Again we note that the lightest DP KK particle is massive due to our choice of the BC, without the need of a dark Higgs vacuum expectation 
value (VEV).

\subsection{Interactions}

\begin{figure}[t]
       \includegraphics[scale=0.43]{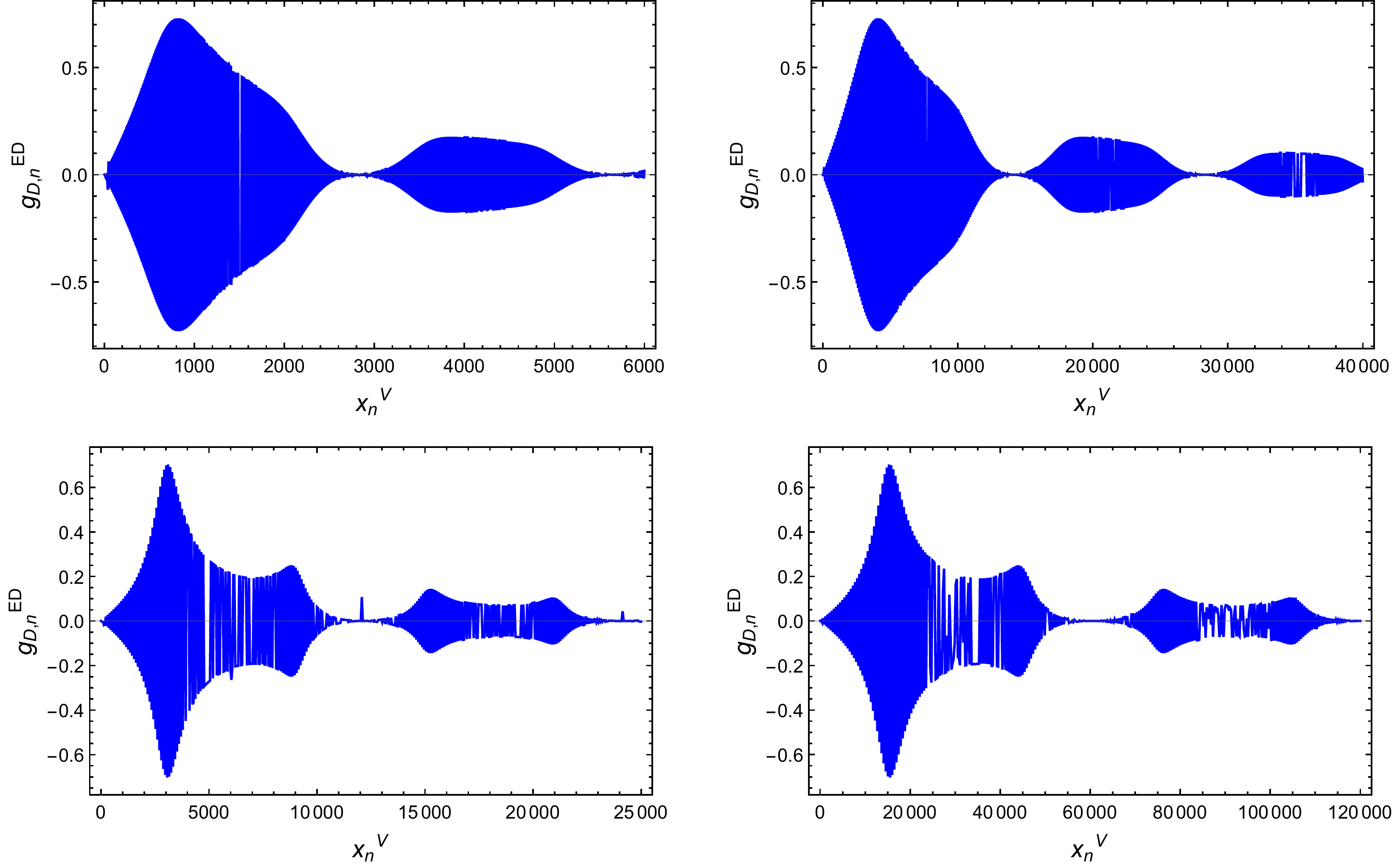}
     
    \caption{Oscillatory behavior of the gauge KK couplings as a function of $x_n^V$, for $ \delta_A=\delta_B=1$, $\alpha=1 $ GeV and $g_{D}=1$ for purposes of demonstration, 
    for BM I (top left), BM II (top left), BM III (bottom left) and BM IV (bottom right).}
    \label{fig:gauge-coupling}
\end{figure}

The couplings between the tower of DP KK states and DM is now simply given by the value of the DP KK tower field wavefunctions evaluated at the DM brane, \textit{i.e.}, the coupling constants 
$g_{D,n}= g_{D} N_n^V/N_1^V  $, where $g_D\equiv g_{5D}N_1^V$ is defined to be the 4-D gauge dark coupling and $N_1^V$ is the normalization of the lowest DP KK state ($n=1$). 
On the other hand, the interaction between the DP and a (zero-mode) SM field, $\psi$, localized `inside' the thick brane is given by the integral 
\begin{equation}
    \int_{\pi r}^{\pi R}dy\, V^{\mu}J_{\mu}\,,
\end{equation}
where $J_\mu$ is the SM current in this version of UED. Since we are interested in the interaction with the conventional SM particles, the zero mode of a generic SM field within the thick 
brane, $\psi$, is simply 
$\psi/\sqrt{\pi L}$, where $(\pi L)^{-1/2}$ is the familiar normalization of the UED fields. The 4-D gauge couplings of the KK gauge fields due to the effect of the thick brane is then 
given by the integral
\begin{align}\label{eq:gauge-coupling-full}
    g_{D,n}^{ED}&\equiv g_{5D}\int_{\pi r}^{\pi R}dy\, \frac{v_{2,n}(y)}{\pi L}\nonumber\\
    &=\frac{g_{D}\,v_{1, n}(\pi r) }{N_1^V\bar{m}_n^V \pi  L }\tan\left(\frac{\bar{m}_n^V \pi  L}{2}\right)\,,
\end{align}
where to obtain the last line we employed the root equation (\ref{eq:rooteq}) above.
For the lower states of the KK tower, since $L\ll R$, we can expand the tangent in the above expression and use the root equation (\ref{eq:rooteq}) in this small mass limit. The result is
\begin{equation}\label{eq:gauge-coupling-small-mass}   g_{D,n}^{ED}\simeq g_{D}\pi\frac{  L }{R}x_n^V\frac{N_n^V}{2N_1^V}\left[\sin(m_n^V \pi r)+\frac{\delta_A}{2}x_n^V\cos(m_N^V \pi r)\right]\,.
\end{equation}
For small values of $x_n^V$ the effect of the extra dimension is thus to reduce the dark coupling by a factor of $\sim L/R$. 
With a compactification scale of size  $R^{-1} \sim 0.1-1 $ GeV and the size of the UED as $L^{-1} \sim 2-10$ TeV, for instance, the interaction with SM particles is thus a factor of 
$\sim 10^{-(3-4)}$ smaller than is the coupling with DM. For more massive particles in the KK tower, the coupling with SM fields is suppressed since $N_n^V \sim (x_n^V)^{-1}$ and thus 
from Eq. (\ref{eq:gauge-coupling-full}) we see that a large $m_n^V$ leads directly to a decrease in the coupling. The full behavior of Eq.(\ref{eq:gauge-coupling-full}) is shown in 
Fig. \ref{fig:gauge-coupling}, where we used the root equation to determine the KK masses. BM I and II (III and IV) produce the same general pattern and whose difference can be seen 
directly from Eq. (\ref{eq:gauge-coupling-small-mass}), although it holds for the full expression. For purposes of demonstration only we have assumed that $g_D=1$ here and in the 
subsequent analysis below. We note that BM II and IV have $L/R$ values ten times smaller than the correspondent ones with $L^{-1}=2$ TeV 
(\textit{i.e.}, BM I and III), thus, in order to compensate for this reduction, ten times as many KK states (via the roots $x_n^V$) are required to match the same behavior.  
Although leading to non-renormalizable interactions in 4-D, it is interesting that the previous result (\textit{i.e.}, the $L/R$ suppression of couplings) generalizes to the case of higher powers of the 
5-D mediator field. Consider a generic 5-D field $\Phi$, then, having interactions with the SM fields that scale as $\Phi^\beta $; this leads to interactions of the form 
$\phi^\beta \int dy \left(v_{2,n}(y)\right)^\beta\sim \phi^\beta \left(\frac{\pi L x_n^V}{R}\right)^\beta$, in the limit of small KK masses for diagonal couplings, where $\phi $ is the lowest KK 
4-D field. 

\begin{figure}[t]
       \includegraphics[scale=0.4]{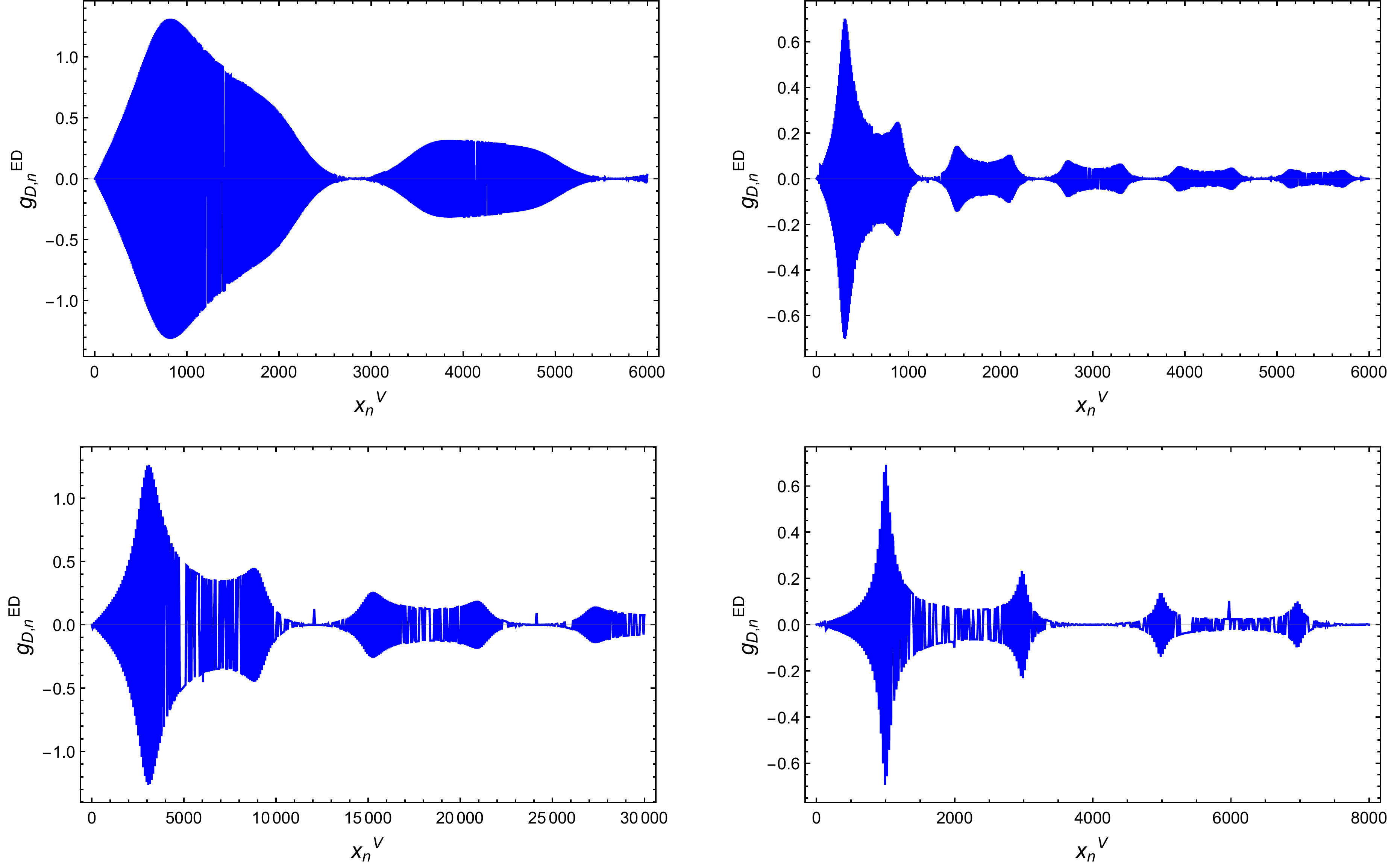}
    \caption{Gauge couplings as a function of $x_n^V$, for $ \delta_A=10 $ (left) or $\delta_B\alpha=10$ GeV (right), while the other parameters are the same as Fig. \ref{fig:gauge-coupling}, for BM I (top) and BM III (bottom).}
    \label{fig:gauge-coupling-delta}
\end{figure}

\begin{figure}
       \includegraphics[scale=0.42]{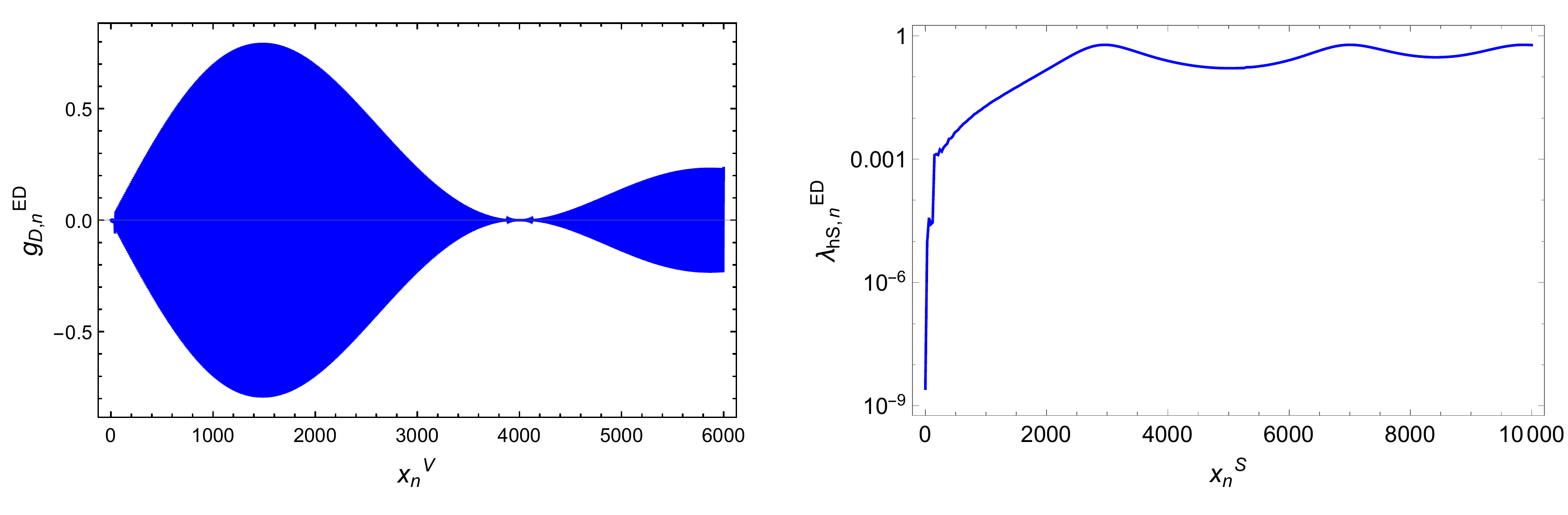}
    \caption{\textit{Left (right):} Gauge (Higgs) couplings as a function of $x_n^V$, for $\delta_B\alpha=10^{-2}$ MeV, while the other parameters are the same as Fig. \ref{fig:gauge-coupling}, for BM I.}
    \label{fig:alpha}
\end{figure}

\begin{figure}
       \includegraphics[scale=0.37]{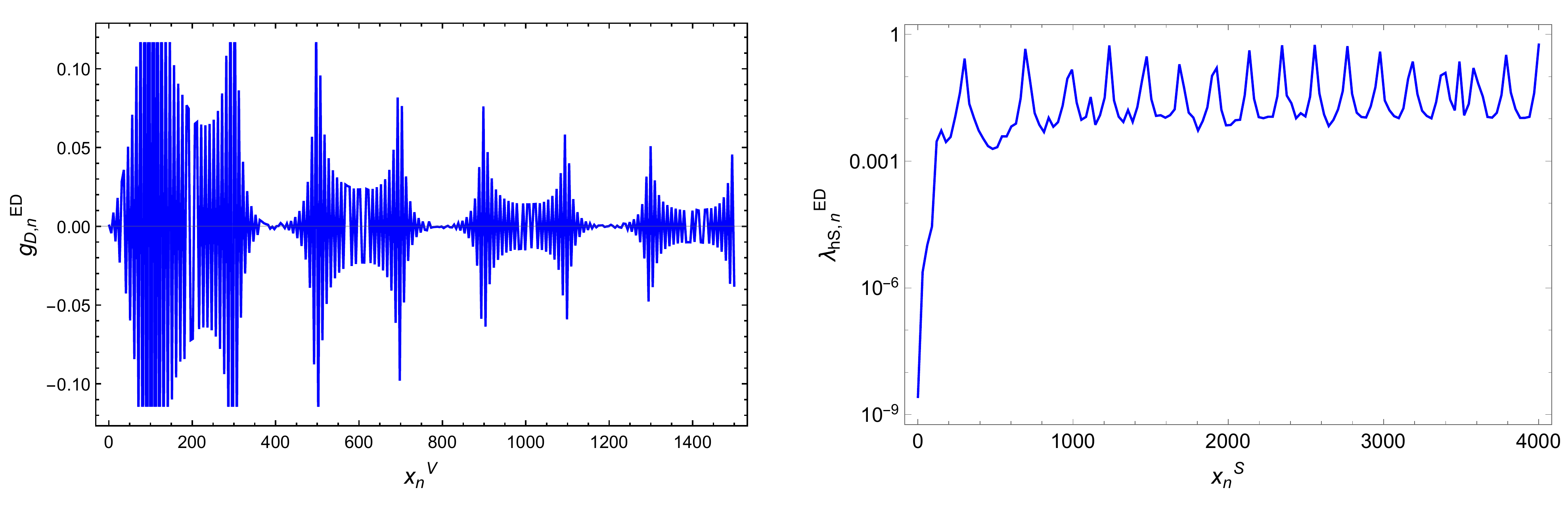}
    \caption{\textit{Left (right):} Gauge (Higgs) couplings as a function of $x_n^V$, for $\delta_B\alpha=10$ TeV, while the other parameters are the same as Fig. \ref{fig:gauge-coupling}, for BM I.}
    \label{fig:alpha-large}
\end{figure}

In Fig. \ref{fig:gauge-coupling-delta} we depict the general behavior of the SM couplings of the gauge KK states for different values of $\delta_A$ and $\delta_B$ in BM I and III; 
these results are for illustrative purposes only. 
Increasing $\delta_A$,  the couplings are also increased as can be seen from Eq. (\ref{eq:gauge-coupling-small-mass}). On the other hand, larger values of $\delta_B\alpha$ lead to an increase 
in the effective masses $\bar{m}_n^V$ thus making the whole oscillatory pattern begin earlier, \textit{i.e.}, the first big peak for BM III is near $x_n^V\sim 1500$ for $\delta_B\alpha=1$ GeV while near 
$x_n^V \sim 500$ for $\delta_B\alpha=10$ GeV.  Fig. \ref{fig:alpha}  presents the KK couplings for BM I with a small value of $\delta_B \alpha$ which is seen to have the opposite behavior, that is, 
it reduces the effective masses $\bar{m}_n^V$ thus leading to a smoother oscillatory pattern which reach maxima at higher values of the roots (compare the BM I in 
Fig. \ref{fig:alpha} with the ones in Figs. \ref{fig:gauge-coupling} and   \ref{fig:gauge-coupling-delta}). As $\delta_B \alpha$ increases the oscillatory pattern becomes sharper with 
oscillations beginning at smaller values of the $x_n^V$. This pattern can be seen in Fig. \ref{fig:alpha-large}, where we considered a large value of $\delta_B\alpha$.

\subsection{Constraints on Dark Matter Couplings}

We now explicitly consider the interactions of the brane localized DM candidate through the DP KK tower to the SM; the DM here is assumed to be a complex scalar field for simplicity. 
The couplings of the DP KK tower with the DM is just $g_{D,n}\equiv g_D N_n^V/N_1^V$, while for the SM it is given by $g_{D,n}^{ED}$ as described in Eq. (\ref{eq:gauge-coupling-full}).

\subsubsection{Direct Search Constraints}
\begin{figure}
    \centering
    \includegraphics[scale=0.35]{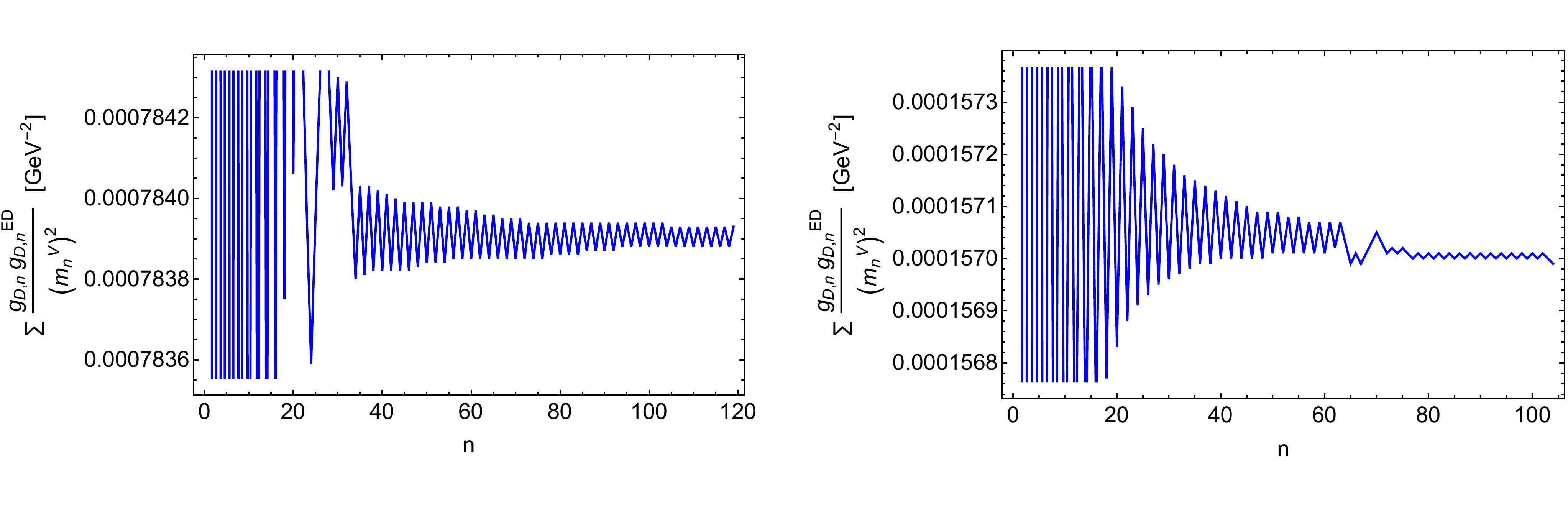}
    \caption{Sum of the first $n$ terms in Eq. (\ref{sigmae}) (the behavior is qualitatively the same for the sum in Eq. (\ref{eq:b})), for BM I (left) and BM II (right). For the the other 
    BM points the pattern is similar. It can be seen that these sums converges quickly, being relatively constant after $n \sim 50$. Small jumps are due to round off errors in the calculation. 
   }
    \label{gauge-sum-DM-relic}
\end{figure}

The mass and couplings of the DM particle are constrained by both direct and indirect experimental searches. In order for the DM not to annihilate into a pair of DP particles (which occurs in 
an s-wave and thus is excluded by Planck results on the CMB \cite{Aghanim:2018eyx} for DM masses in our range of interest), it should be lighter than the lightest DP KK state. Thus, the 
DM mass must be smaller than $x_1^V/R$, where $x_1^V$ is the lowest KK root, which generally lies in the range $\sim 0.2-0.5$ for the BMs considered here. The mass scale for the DM 
is, therefore, set by both the compactification radius $R$ and the value of the first root obtained from the DP KK tower mass eigenvalue equation above. 
Due to the magnitude of the DM mass ($\sim 20-400$ MeV, as we will see below) and the corresponding small recoil energies in direct detection, DM scattering off electrons provides greater sensitivity \cite{Battaglieri:2017aum}; the scattering cross section in this case is \cite{Essig:2011nj,Berlin:2014tja,Emken:2017erx}
\begin{equation}\label{sigmae}
    \sigma_e=\frac{\mu^2}{4   \pi  }\left(\sum_n\frac{g_{D,n} g_{D,n}^{ED}}{(m_n^V)^2}\right)^2\,,
\end{equation}
where a form factor of unity has been assumed and the reduced mass $\mu=m_em_{DM}/(m_e+m_{DM})\sim m_e$, since $m_{DM}^2\gg m_e^2$. Given the couplings and masses from 
the above considerations the KK sum that appears in this expression converges quite rapidly with only the first few terms being numerically significant (see Fig. \ref{gauge-sum-DM-relic}). 
The resulting scattering cross section has 
been/can be constrained using the results from XENON10 \cite{Angle:2011th}, XENON100 \cite{Aprile:2016wwo}, DarkSide-50 \cite{Agnes:2018oej} or SENSEI \cite{Crisler:2018gci}. 
The corresponding values of the elastic scattering cross sections for the four different BM points are shown in Table \ref{tab:relicDens}.

\subsubsection{Indirect constraints}
 
Considering the standard scenario of thermal freeze-out, given the DM mass range of interest, the resulting final states from DM pair annihilation can be $e^+e^-$ and $\mu^+\mu^-$,   as well as three generations of (essentially) massless neutrinos since the DP couples to $B-L$.  Note that as mesons do not have either baryon nor lepton number, a pair of pions is not an 
accessible channel of annihilation. The cross section away from a KK resonance can be approximately expanded in powers of the relative DM annihilation, 
$v^2$, and is given by $\sigma v \approx a+bv^2$, where, for a vector mediator and a complex scalar DM, the annihilation cross section is dominantly a p-wave with coefficients $a=0$ 
and for a particular final state fermion $f$  \cite{Berlin:2014tja}
\begin{equation}\label{eq:b}
   b_f=\frac{m_{DM}^2}{ 6 \pi }\sqrt{1-\frac{m_f^2}{m_{DM}^2}}\left(1-\frac{m_f^2}{2m_{DM}^2}\right)\left(\sum_n\frac{g_{D,n} g_{D,n}^{ED}}{(m_n^V)^2-4m_{DM}^2}\right)^2\,,
\end{equation}
where $m_f$ is the mass of the final states.  The sum above converges rapidly (see Fig. \ref{gauge-sum-DM-relic}) because the couplings $g_{D,n}^{ED}$ go to zero rather quickly 
for higher roots $x_n^V$ in addition to the presence of the DP tower propagator masses which rapidly decreases the possible contribution of heavier DP states.

We employ the following expression to determine the value of the coupling $g_D$ which gives the observed value of relic density of DM \cite{Berlin:2014tja} 
\begin{equation}
    \Omega h^2 \simeq \frac{x_f\, 1.07\times 10^9  \,\text{GeV}^{-1}}{g_*^{1/2}m_{Pl}(a+3b/x_f)}\,,
\end{equation}
where $m_{Pl}$ is the Planck mass and $x_f\equiv m_{DM}/T_f$ is the usual ratio between the DM mass and the temperature at the freeze-out, which is taken to be $x_f=20$. 
The effective number of degrees of freedom for the range of DM masses of interest here ($20-400$ MeV, as we shall see) is $g_*\simeq 10.75$, since the temperature at the freeze-out 
is $\sim 1-20$ MeV. Setting $\Omega h^2=0.12$ \cite{Aghanim:2018eyx} our results for $g_D$ 
are found in Table \ref{tab:relicDens}. The lightest DP KK mass is shown here together with the value for the DM mass. A 
slightly heavier DM particle can be considered, provided that $m_{DM}<m_{1}^V$ remains valid, although the results are practically unchanged. Smaller values of $m_{DM}$ increase 
the value of $g_D$ needed to obtain the observed DM relic abundance, as seen in Eq. (\ref{eq:b}). Recall that $g_D$ is defined here to be the coupling of the lowest DP KK tower state to 
the DM and that all of the values of $\alpha_D=g^2_D/4\pi$ shown here are $<0.4$.

\begin{table}[h] 
       \centering 
           \begin{tabular}{c| c c c c}
           \hline\hline
    BM  &   I &  II &  III &  IV\\
   \hline
   $\delta_A=1$ & & & &\\
   \hline
    $m_{1}^V$ [MeV] &430 &  430 & 43 & 43  \\
   $m_{DM}$ [MeV] &400  & 400 & 40 & 40  \\
   $g_D$  &0.89 & 1.98 & 0.96 &2.15 \\
         $\sigma_{e} $ [cm$^{2}$] & $1.1\times 10^{-40}$&  $1.1\times 10^{-40}$ & $1.6\times 10^{-38}$ & $1.6\times 10^{-38}$  \\
         \hline
          $\delta_A=1/2$ & & & &\\
          \hline
           $m_{1}^V$ [MeV] &460 &  460 & 46 & 46  \\
   $m_{DM}$ [MeV] &430  & 430 & 43 & 43  \\
   $g_D$  &0.88 & 2.00 & 0.95 &2.12 \\
         $\sigma_{e} $ [cm$^{2}$] & $8.8\times 10^{-41}$&  $9.6\times 10^{-41}$ & $1.2\times 10^{-38}$ & $1.2\times 10^{-38}$  \\
         \hline
         $\delta_A=10$ & & & &\\
          \hline
            $m_{1}^V$ [MeV] &230 &  230 & 23 & 23  \\
   $m_{DM}$ [MeV] &200  & 200 & 20 & 20  \\
   $g_D$  &0.88 & 1.95   & 0.93 &2.08 \\
        $\sigma_{e} $ [cm$^{2}$] & $1.1\times 10^{-39}$&  $1.1\times 10^{-39}$ & $1.4\times 10^{-37}$ & $1.4\times 10^{-37}$  \\
             \hline
         $\delta_B\alpha=1/2$ GeV & & & &\\
          \hline
            $m_{1}^V$ [MeV] &430 &  430 & 43 & 43  \\
   $m_{DM}$ [MeV] &400  & 400 & 40 & 40  \\
   $g_D$  &0.81 & 1.98  & 0.96 &2.14 \\
        $\sigma_{e} $ [cm$^{2}$] & $7.7\times 10^{-41}$&  $1.1\times 10^{-40}$ & $1.5\times 10^{-38}$ & $1.5\times 10^{-38}$  \\
             \hline
         $\delta_B\alpha=10$ GeV & & & &\\
          \hline
            $m_{1}^V$ [MeV] &430 &  430 & 43 & 43  \\
   $m_{DM}$ [MeV] &400  & 400 & 40 & 40  \\
   $g_D$  &0.81 & 1.98  & 0.96 &2.14 \\
        $\sigma_{e} $ [cm$^{2}$] & $7.8\times 10^{-41}$&  $1.1\times 10^{-40}$ & $1.5\times 10^{-38}$ & $1.5\times 10^{-38}$  \\
           \hline
         $\delta_B\alpha=10$ TeV & & & &\\
          \hline
            $m_{1}^V$ [MeV] &430 &  430 & 43 & 43  \\
   $m_{DM}$ [MeV] &400  & 400 & 40 & 40  \\
   $g_D$  &0.72 & 1.98  & 0.94 &2.14 \\
        $\sigma_{e} $ [cm$^{2}$] & $3.0\times 10^{-41}$&  $1.1\times 10^{-40}$ & $1.3\times 10^{-38}$ & $1.5\times 10^{-38}$  \\
         \hline\hline \end{tabular}
 \caption{ 4-D dark coupling $g_D$ needed to obtain the observed DM relic density, with $ \delta_A=1/2,1 \text{ or } 10$ (with $\delta_B\alpha=1$  GeV), and $\delta_B\alpha=1/2, 10$ GeV  or $10$ TeV (with $\delta_A=1$).  The maximum value for the DM  mass and the lightest DP mass are also shown. For BM II and IV the DM particle is lighter than muons, thus the only channels accessible is $e^+e^-$ and neutrinos. The 
 cross section for the scattering off of an electron by a DM particle is shown in the last row, using the same gauge coupling needed to satisfy the DM relic abundance.}   \label{tab:relicDens}
\end{table}

The sum appearing in Eq. (\ref{eq:b}) is qualitatively the same for BM I and III (or BM II and IV).  This comes from the fact that the differences in the coupling $g_{D,n}^{ED}$ (roughly one 
order of magnitude, and whose behavior are very well described by Eq. (\ref{eq:gauge-coupling-small-mass})) is compensated by presence of the different DP masses in the denominator of 
Eq. (\ref{eq:b}). Although the DM masses are one order of magnitude smaller for BM III (IV) than for BM I (II), leading apparently to smaller cross sections, this reduction is compensated 
for by the normalization of the first root, \textit{i.e.}, $N_1^V$, which is roughly three times larger for BM I (II) than in BM  III (IV) (see Fig. \ref{fig:nu-plot}).
The most `promising' among these BM appears to be BM I and III, since even with smaller DM masses, which in turn leads to larger couplings. Increasing or decreasing the thick 
brane parameter $\delta_B\alpha$ (\textit{e.g.}, $\delta_B\alpha=1/2$ GeV or $\delta_B\alpha=10$ GeV) does not significantly change these numerical results. A large increase of the combination $\delta_B\alpha$ ($\delta_B\alpha=10 $ TeV) has also no significant impact on the results presented, being the most relevant change for BM I, in which the scattering cross section is decreased by almost one order of magnitude.  Variations in the parameter $\delta_A$, on the other hand, 
do lead to some numerical modifications here. If $\delta_A=1/2$ is assumed, the numerics are unchanged but if instead one chooses 
$\delta_A=10$, the results are slightly modified; we gather all of these variations in Table \ref{tab:relicDens}. Note that the values of the couplings $g_D$ are slightly reduced when 
$\delta_A=10$, however, the scattering cross section $\sigma_e$ in this case is found to be roughly one order of magnitude smaller, thus BM III and IV are essentially excluded employing 
the current direct search experimental results \cite{Angle:2011th,Aprile:2016wwo,Agnes:2018oej,Crisler:2018gci}.

\section{Scalar Mediator Fields in the Bulk}\label{sec:scalar}

We now consider instead a scalar field $S$ in the bulk and investigate whether the ED can suppress the corresponding coupling with SM fields as we saw above in the case of a bulk 
gauge field. A real scalar field without an additional $\mathbb{Z}_2$ symmetry would give rise to potentially dangerous mass mixing terms between the scalar and the SM Higgs, thus 
spoiling the agreement of the Higgs couplings measurements with SM expectations. Therefore the imposition of a $\mathbb{Z}_2$ symmetry would be required. However, a real scalar 
with an exact $Z_2$ symmetry would be stable and could be a possible DM candidate. Since this particle is not assumed to be DM here, it must decay into, \textit{e.g.}, our DM candidate. These 
issues can be avoided if the scalar field is complex with a Higgs-like potential, couples to the DP and obtain a VEV. This VEV, in turn, gives a mass to the DP in the bulk, but this induces 
only order one changes in the results obtained in the previous sections. 

However, for simplicity  we will consider the following mechanism for a real scalar field in the bulk with a $\mathbb{Z}_2$ symmetry, without loss of generality. The action for the scalar field in the bulk is (see footnote in the gauge field case) 
\begin{align}\label{eq:actionS}
    S=\int d^4x\int_0^{\pi R} dy& \left[ \frac{1}{2}\partial_AS\partial^AS+\frac{1}{2}\partial_\mu S\partial^\mu S\cdot \delta_SR\, \delta(y)+\frac{1}{2}\partial_\mu S\partial^\mu S\cdot \delta_T R\, \theta(y)\right]\nonumber\\
    &+ \int d^4x\int_{\pi r}^{\pi R} dy\,\lambda_{5hS}S^2 |H|^2\,,
\end{align}
where $H$ is the SM Higgs field and as before we have added BLKTs and the step-function $\theta(y)$ as given by Eq. (\ref{eq:theta}).  In principle the scalar field may have a bulk mass as 
well as quartic potential terms, however, for simplicity, we will ignore this possibility as they will not play any essential role in what follows. 

We then expand the 5-D scalar field in a KK tower of states and the equation of motion is found to be similar to the one for the gauge field above. However, now we have a bulk mass term 
but only for the region within the thick brane, \textit{i.e.}, $\pi r< y\leq \pi R$,  given by $m_b^2\equiv\lambda_{5hS}v_h^2/(\pi L)$, arising from the SM Higgs VEV. Note the presence of the square 
of the factor  $(\pi L)^{-1/2}$ being the normalization of the zeroth mode of SM Higgs field as in UED and here $v_h$ is the Higgs VEV. 

The physical masses are determined expanding the 5-D scalar field in the action  (\ref{eq:actionS}) in a KK tower of states. The masses will have contributions due to the fields in the two distinct regions in the ED, \textit{i.e.}, in the region $R>y>r$ and $r>y>0$.  The physical KK masses that appears in the 4-D action are thus given by
\begin{align}
   (m_{p,n}^S)^2\delta_{m,n}&\equiv \int_0^{\pi r}dy\, \partial_ys_{1,m}(y)\partial_ys_{1,n}(y)+ \int_{\pi r}^{\pi R}dy\, \partial_ys_{2,m}(y)\partial_ys_{2,n}(y)\nonumber\\&+m_b^2\int_{\pi r}^{\pi R}dy\, s_{2,m}(y)s_{2,n}(y)\nonumber\\
   &= (m_n^S)^2+\lambda_{hS,m,\,n}^{ED}v_h^2\,,
\end{align}
where the first integrals yield the quantity $(m_n^S)^2$ after integrating by parts and we defined the 4-D couplings $\lambda_{hS,m,\,n}^{ED}$ as
\begin{equation}\label{eq:lambdaED}
    \lambda_{hS,m,\,n}^{ED}\equiv  \lambda_{5hS}\int_{\pi r}^{\pi R}dy\, \frac{s_{2,m}(y)s_{2,n}(y)}{\pi L}\,. 
    \end{equation}
The equation of motion for $s_1(y)$ is found to be 
\begin{equation}\label{eq:eomS1}
  \partial_y^2s_{1,n}+(m_n^S)^2s_{1,n}+(m_n^S)^2\delta_S R \delta(y)  s_{1,n}=0\,,
\end{equation}
whose solution is the same as Eq.(\ref{eq:v1}), with the trivial replacement $\delta_A\rightarrow \delta_S$.
For $\pi r<y\leq \pi R$, however, the equation of motion for $s_2(y)$ is now 
\begin{equation}\label{eq:eomS}
  \partial_y^2s_{2,n}+(m_n^S)^2s_{2,n}+m_b^2s_{2,n}+(m_n^S)^2\delta_T\alpha R  s_{2,n}=0\,,
\end{equation}
where we may define an effective mass as
\begin{equation}\label{massbarS}
    (\bar{m}_n^S)^2=\frac{(x_n^S)^2}{R^2}(1+ \delta_T \alpha R) +m_b^2\,.
\end{equation}
Using the above definition, the solution for $s_{2,n}$, the normalization $N_n^S$ and the root equation have the same structure of Eqs. (\ref{eq:v2}), (\ref{eq:Nu}) and  (\ref{eq:rooteq}), respectively, with the replacements $\delta_A\rightarrow \delta_S$, $\delta_B\rightarrow\delta_T$, $\bar{m}_n^V\rightarrow\bar{m}_n^S$ and $x_n^V\rightarrow x_n^S$. Of course, the solutions of the root equation are quantitatively different from the gauge field case. Once the quantized masses $m_n^S$ and the 4-D $\lambda_{hS,m,\,n}^{ED}$ are found through the root equation, the physical masses are straightforwardly determined.

\subsection{The Higgs portal}

The coupling between the SM Higgs and scalar mediator field $S$ is given by the Higgs portal interaction, $\lambda_{5hS}|H|^2 S^2$. We split the  effect of the extra dimension on 
the coupling into two different contributions: that arising from the particles at the same KK level `$n$' ($m_m^S=m_n^S$) and for that contributions arising 
between different levels of the KK tower ($m_m^S\neq m_n^S$). These are seen to arise from the integral over the ED in Eq. (\ref{eq:lambdaED}).
The integral for $m=n$, after applying the root equation, is given by
\begin{align}\label{intm=n}
 \lambda_{hS,n}^{ED}=&\frac{\lambda_{hS}\,s_{1,n}^2}{2(N_1^S)^2}\left[\csc^2(\bar{m}_n^S\pi L)- \frac{1 }{  \bar{m}_n^S \pi L}\cot(\bar{m}_n^S\pi L)\right]\,,
\end{align}
while the integral for $m\neq n$ gives
\begin{align}
  \lambda_{hS,m,\, n}^{ED}=& \frac{\lambda_{hS}\,s_{1,m}s_{1,n}}{(N_1^S)^2\pi L}\left[\frac{\bar{m}_n^S\cot(\bar{m}_n^S \pi L)-\bar{m}_m^S\cot(\bar{m}_m^S \pi L) }{(\bar{m}_m^S)^2-(\bar{m}_n^S)^2}\right] \,,
\end{align}
and where $s_{1,m(n)}$ are here understood to be evaluated at $y=\pi r$. As before, we may define a 4-D quartic coupling as $\lambda_{hS}\equiv \lambda_{5hS}(N_1^S)^2$. Since the 
bulk scalar mass is not small, the effective mass $\bar{m}_m^S$ is not as well, therefore we cannot expand the trigonometric functions in the above expressions, as we have done in the gauge field case.

\begin{figure}
\centering
 \includegraphics[scale=0.4]{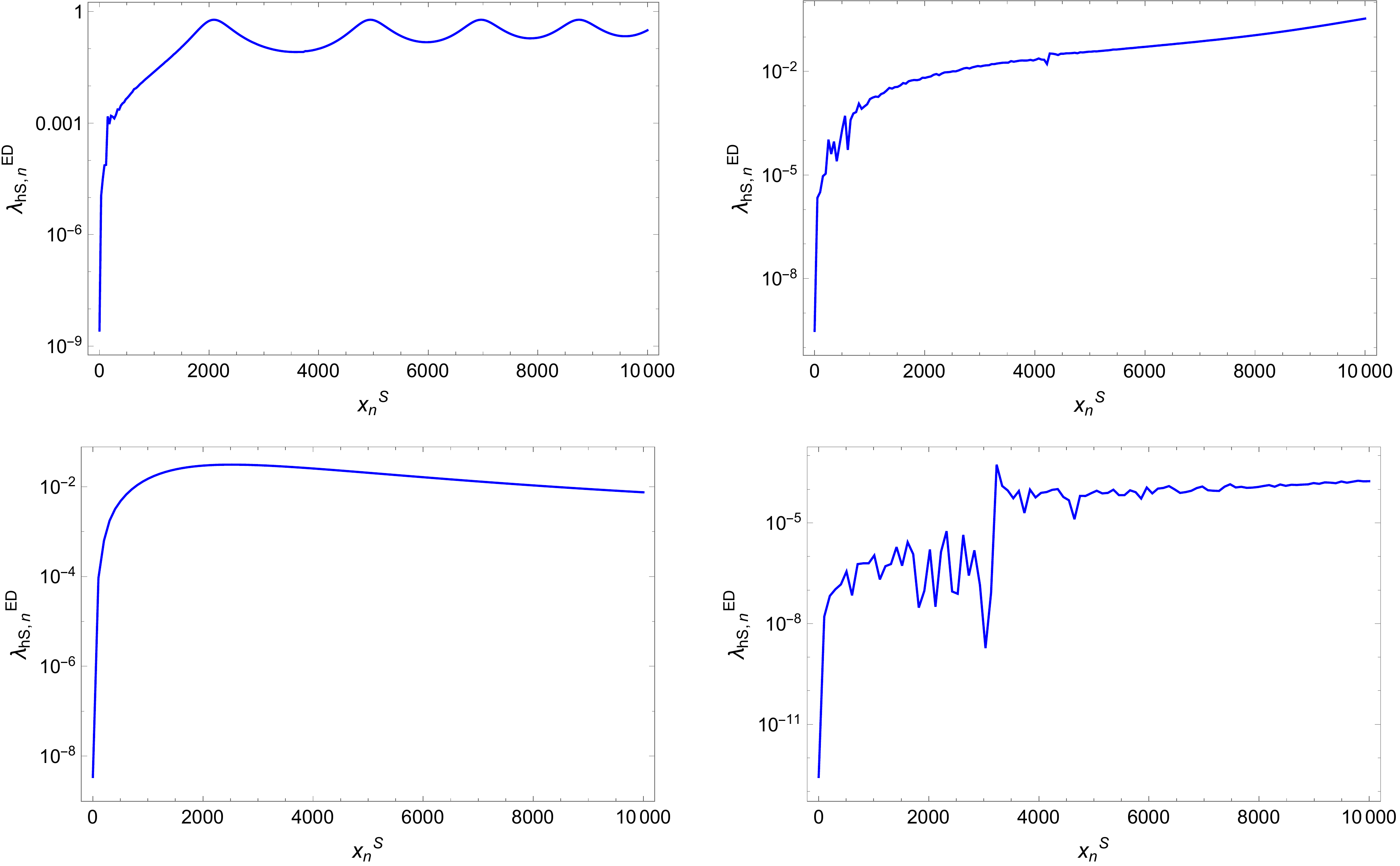}
\caption{Higgs portal couplings as a function of $x_n^S$, for $ \delta_S=\delta_T=1$, $\alpha= 1$ GeV and $\lambda_{hS}=1$,  for BM I (top left), BM II (top left), BM III (bottom left) and
 BM IV (bottom right) as described above.}
 \label{fig:higgs-portal-coupling}
\end{figure}

\begin{figure}
\centering
 \includegraphics[scale=0.48]{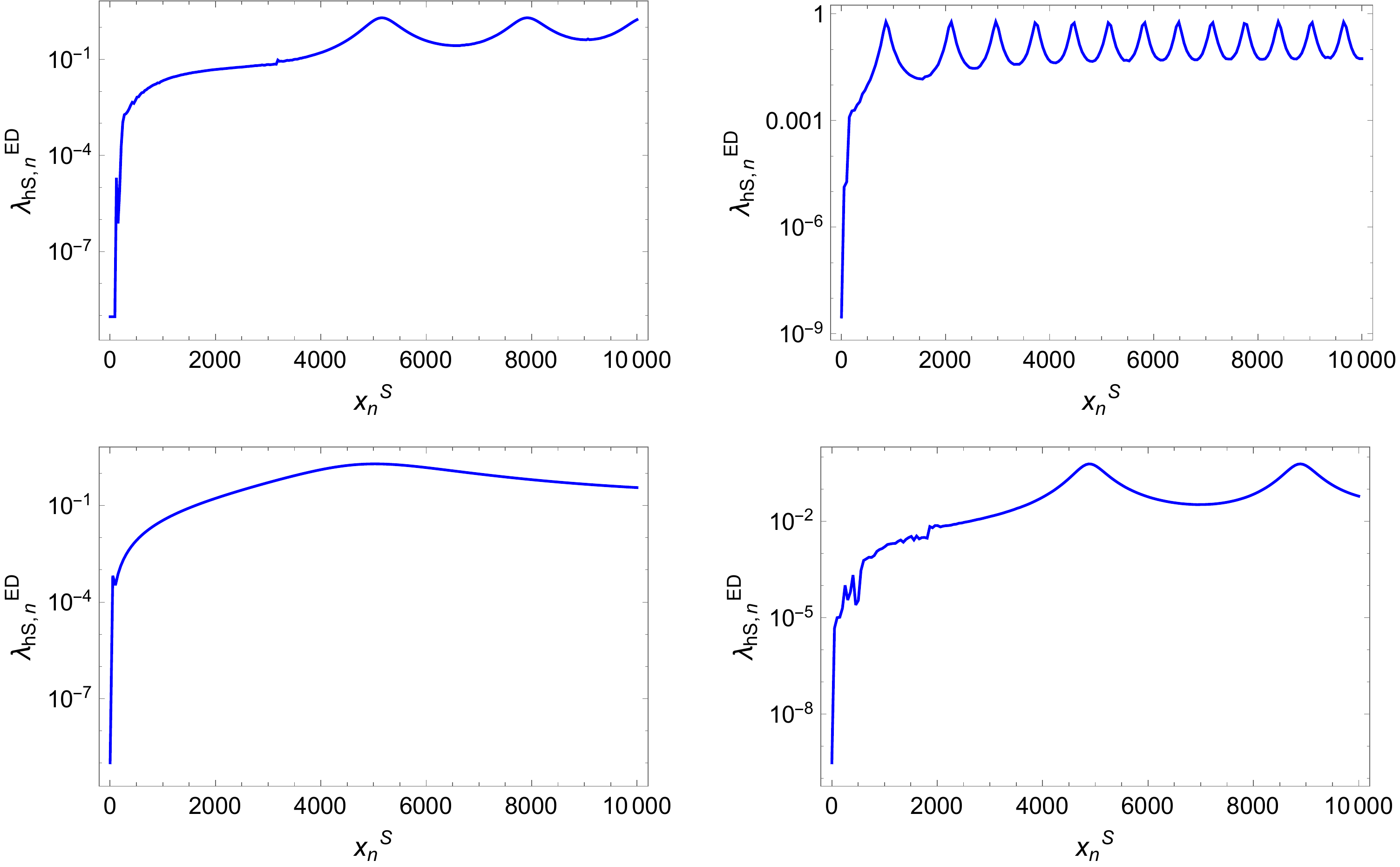}
\caption{Higgs portal couplings as a function of $x_n^S$, for $ \delta_S=10$ (left) or $\delta_T\alpha=10$ GeV (right), but with the other parameters unchanged, for BM I (top) and BM II (bottom).}
\label{fig:higgs-portal-couplingDelta}
\end{figure}

\begin{figure}
    \centering
    \includegraphics[scale=0.35]{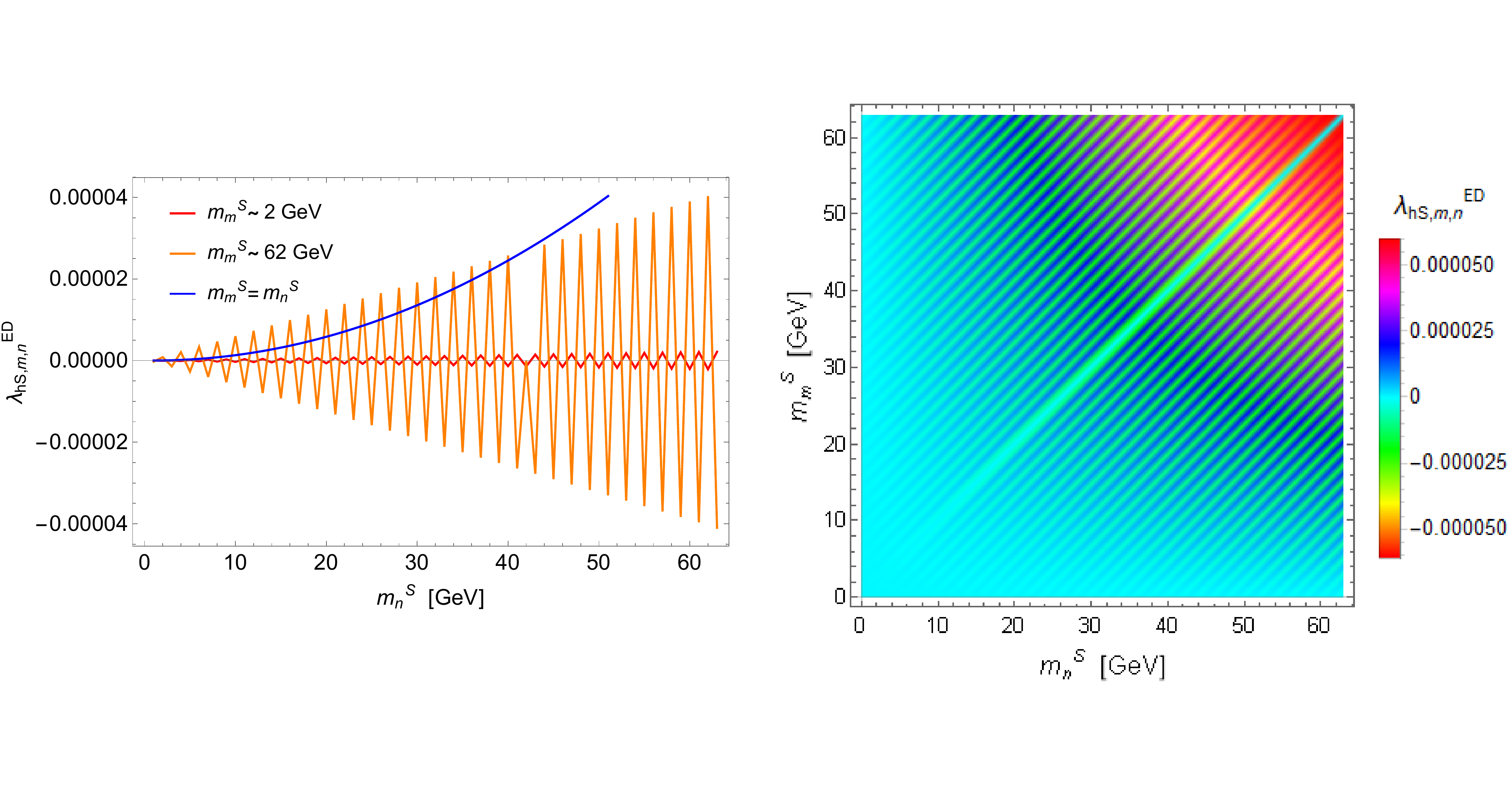}
    \caption{\textit{Left:} The coupling $\lambda_{hS,m,n}^{ED}$, for BM I, with $m$ fixed for different values of $m_m^S$ and $\lambda_{hS}=1$. \textit{Right:} Coupling $\lambda_{hS,m,n}^{ED}$ up to $m_{m(n)}^S=62$ GeV. }
    \label{fig:higgs-portal-upt-62}
\end{figure}

The numerical results obtained from Eq. (\ref{intm=n}) are depicted in Fig. \ref{fig:higgs-portal-coupling}, where the pattern of BM I and II (or III and IV) are explained by the  same reasoning  
as described for the gauge interaction in the last section. Different values of $\delta_S$ and $\delta_T\alpha$ are shown in Figs. \ref{fig:alpha} and \ref{fig:higgs-portal-couplingDelta}.
The behavior of the off-diagonal couplings can be  seen in Fig.~\ref{fig:higgs-portal-upt-62}. The couplings here are seen to oscillates around zero, but are generally increasing in absolute 
value for heavier masses. 

\subsubsection{Higgs Decay Constraints}

The lower members of the bulk scalar KK tower are generally lighter than the Higgs boson itself, so that the Higgs can decay into a pair of potentially neutral long-lived particles of the 
KK tower. The most recent constraint for the Higgs decay into invisible particles is given by the ATLAS experiment \cite{Aaboud:2018sfi}.
The $h\rightarrow S_nS_m$ decay rate can be calculated employing the standard result \cite{Tanabashi:2018oca}
\begin{equation}
    d \Gamma_{m,n}=\frac{1}{32 \pi^2} |\mathcal{M}|^2\frac{|\mathbf{p}_{m,n}|}{m_h^2}d\Omega\,,
\end{equation}
where $m_h= 125$ GeV is the SM Higgs mass, the outgoing momentum is
\begin{equation}
    |\mathbf{p}_{m,n}|=|\mathbf{p}_{n,m}|=\frac{[(m_h^2-(m_m+m_n)^2)(m_h^2-(m_m-m_n)^2)]^{1/2}}{2m_h^2}\,,
\end{equation}
and the integral over the solid angle is $4\pi$ for $m\neq n$ and $2\pi$ for $m=n$. The invariant amplitude $|\mathcal{M}|$ is $2\lambda_n^{ED}v_h$ for $m=n$ and 
$\lambda_{m,n}^{ED}v_h$ for $m\neq n$, where $v_h$ is the Higgs VEV. Here we take $\lambda_{hS}=1$ for simplicity. Then, the total decay rate due to the tower of kinematically 
accessible scalar KK particles is given by $\Gamma^{ED}=\sum_{m,n} \Gamma_{m,n}$. When $m=n$ the Higgs can decay into particle pairs with individual masses up to $\sim 62.5$ 
GeV while when $m\neq n$, the masses of the outgoing KK particles can range up to $\simeq m_h$.
The  branching fractions for this invisible decay $\mathcal{B}(h\rightarrow S_m+S_n)= \Gamma^{ED}/(\Gamma_{H}^{SM}+\Gamma^{ED})$, where  $\Gamma_{H}^{SM}\simeq 4$ MeV is 
the SM Higgs total width \cite{Tanabashi:2018oca}, summed over kinematically accessible KK states are shown in Table \ref{tab:higgsdecay} for the four BM models described above 
in Table \ref{table:param}.

\begin{table}[h] 
       \centering 
           \begin{tabular}{c| c c c c}
           \hline\hline
    BM  &   I &  II &  III &  IV\\
   \hline
   
   $\mathcal{B}(h\rightarrow S_m+S_n)$  &$3.6 \times 10^{-3}$  & $3.7 \times 10^{-5}$  & $0.87$ & $3.5 \times 10^{-7}$  \\
         \hline\hline \end{tabular}
 \caption{ Branching fractions for the invisible Higgs decay into two KK scalars summed over all kinematically allowed states with  $ \delta_S=\delta_T=1$, $\alpha= 1$ GeV and 
 $\lambda_{hS}=1$.}   \label{tab:higgsdecay}
\end{table}

The differences between the branching fractions for the various BM arise from two distinct factors: $i)$ the difference in the quartic coupling, as seen, \textit{e.g.},  in the BMs with 
$L^{-1}=10$ TeV (BM II and IV), which are two or three orders of magnitude smaller than the ones where  $L^{-1}=2$ TeV; and $ii)$ the total number of KK states that need to be summed 
over which can contribute to Higgs decay. If $R^{-1} =1$ GeV (BM I and II) there are $\sim 62$ states with masses below 62.5 GeV, while for $R^{-1} =100$ MeV (BM III and IV) there are 
$\sim 10$ times as many. This fact is seen in the final numerics (BM III has a larger decay rate than BM I) once there are more KK tower states that contribute to the sum. The expected  
larger decay rate obtained for BM IV in comparison to BM II (given $ii$ above) is compensated by the smaller couplings that appear in this case (by roughly three orders of magnitude), 
thus leading to an overall smaller branching fraction instead of a potentially expected larger one. Clearly BM III is disfavored by these results.

\begin{figure}
    \centering
    \includegraphics[scale=0.36]{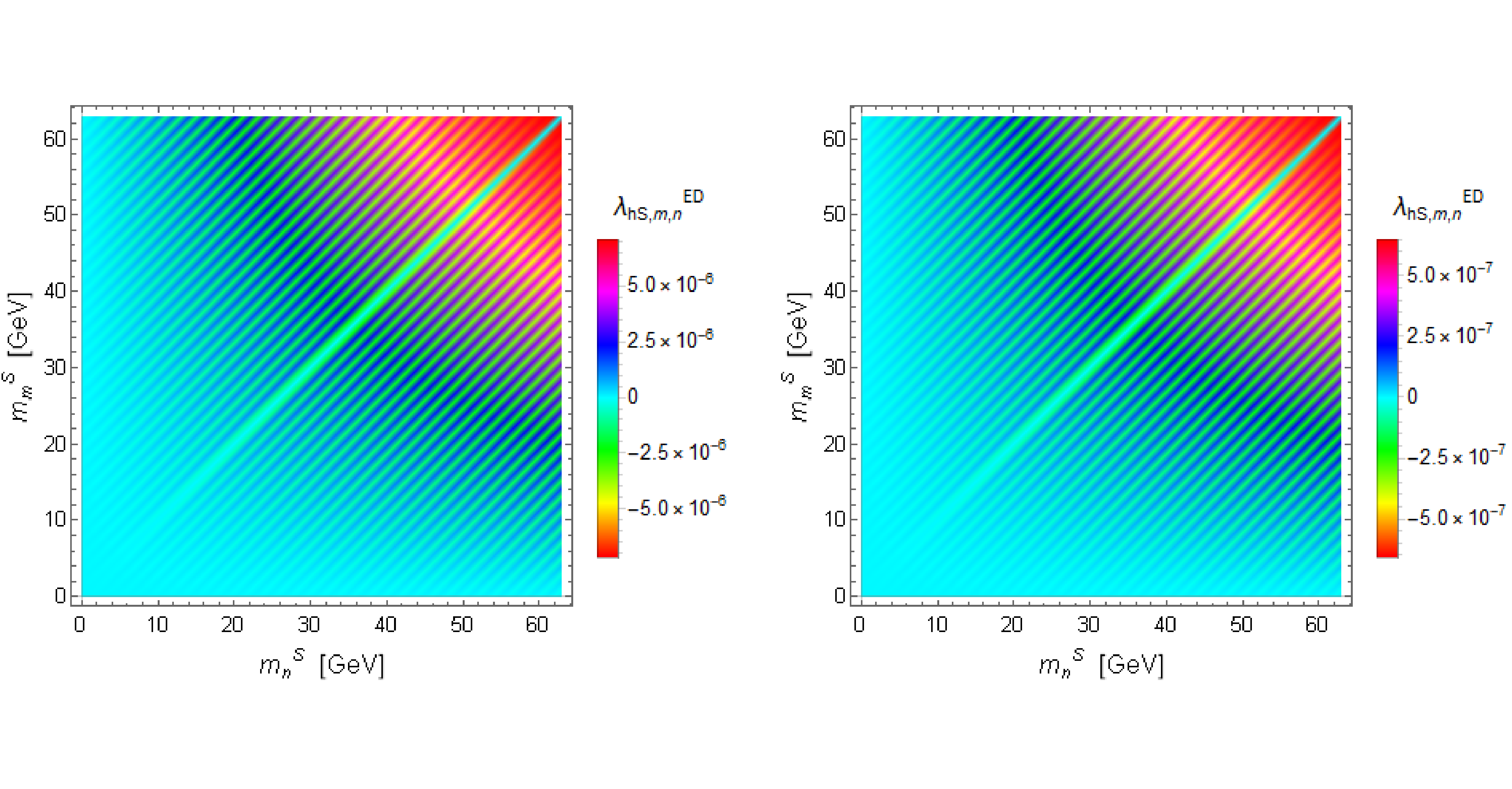}
    \caption{Couplings $\lambda_{hS,m,n}^{ED}$ up to $m_{m(n)}^S=62$ GeV for BM II, with $\delta_T\alpha=1$ GeV (left) and $\delta_T\alpha=10$ GeV (right). The other parameters are the same as in Fig. \ref{fig:higgs-portal-coupling}.}
    \label{fig:higgs-portal-upt-62delta}
\end{figure}

In order to elucidate the role of the BLKT in the previous results, we consider BM II as an example and analyze the normalized decay rate for different values of both $\delta_S$ and 
$\delta_T\alpha$.  For $\delta_S=10$, $  \mathcal{B}(h\rightarrow S_m+S_n)=1.4 \times 10^{-2}$, summing over all kinematically accessible KK states, while for $\delta_S=1/2$ we obtain 
instead $\mathcal{B}(h\rightarrow S_m+S_n)=7.0 \times 10^{-5}$. Taking $\delta_T\alpha=10$ GeV, \textit{e.g.}, the corresponding result is $\mathcal{B}(h\rightarrow S_m+S_n)=3.4 \times 10^{-7}$, while for $\delta_T\alpha=10$ TeV the branching ratio is $\mathcal{B}(h\rightarrow S_m+S_n)=2.8 \times 10^{-9}$. 
These differences in the branching fractions arise mainly due to the order of magnitude of the effective Higgs portal couplings, which are an order smaller for $\delta_S=1$ than 
for $\delta_S=10$ (see the left bottom panel of Fig. \ref{fig:higgs-portal-couplingDelta} for $m=n$, though the two result are similar for $m\neq n$). For larger $\delta_T\alpha $ the couplings 
are similar when $m=n$ (compare the right top panel of Fig. \ref{fig:higgs-portal-coupling} with the right bottom panel of Fig. \ref{fig:higgs-portal-couplingDelta}), but are also 
one order of magnitude smaller when $m\neq n$, as can be seen in Fig. \ref{fig:higgs-portal-upt-62delta}, thus leading to a smaller branching fraction, since the couplings for $m=n$ 
are in general smaller than the ones for $m\neq n$. 
From all these previous results, one sees that all benchmark models but BM III are far below the LHC constraints on the invisible Higgs decay 
$\mathcal{B}(h\rightarrow inv)\lesssim 0.2$ \cite{Aad:2015pla,Duerr:2015aka}. In this scenario, because the scalar field couples to DM through a Higgs portal-like coupling, the 
annihilation process proceeds only through loops and so its contribution is negligible in the determination of the DM relic abundance.

\section{Discussion and Conclusions}\label{sec:discussion}

In this paper we have presented a mechanism that naturally suppresses the various dark couplings with the SM which occur through a bulk mediator by employing a flat, single ED. The 
SM is confined in a thick brane while the DM is  localized on the other (thin) brane at the opposite end of the interval. For both vector and scalar mediator fields in the bulk the mechanism 
is qualitatively similar: the coupling with the SM is significantly reduced in comparison to the corresponding coupling with DM, which we explicitly examined for both the gauge and the Higgs 
portal interactions. The sizes of the couplings in turn depend upon the ratio of the 5-D compactification radius, $R^{-1} \sim 0.1-1$ GeV, and the SM brane thickness, 
$L^{-1} \sim 2-10$ TeV, whose range was chosen to avoid LHC and other experimental constraints. In this scenario, a kinetic mixing between the DP and the SM hypercharge gauge field 
is not needed once the DM carries a dark charge and the SM particles carry 
a $B-L$ charge. This special feature arises from a BLKT within the thick brane since the wave function in that small region is different from the one in the rest of the ED. 

The observed DM relic abundance was obtained in the four BM considered but whose annihilation channels and rates clearly rely on the DM mass. Since the DM mass is assumed to be smaller than that of the lightest DP KK state, its value is bounded by the compactification radius $R$ and the smallest DP mass/root $x_1^V$. 

The entire approach we considered here relies upon the use of BLKTs in order to obtain different mass parameters ($m_n^{V(S)}$ and $\bar{m}_n^{V(S)}$) in the two regions of ED (the thick brane region, and the rest of the ED).
However, if only a scalar mediator field was present in the bulk, the interaction with the Higgs would still induce an effective mass (\ref{massbarS}) different from $m_n^S$, leading to distinct wave functions in the two regions of the ED. Therefore, in this hypothetical situation, the resulting difference in the wave functions in these two regions could arise solely due to the 
Higgs VEV, therefore not necessarily requiring the BLKT. Even so, as presented before, a gauge field in the bulk does require a BLKT in order to suppress the couplings with SM.

Direct detection experiments employing DM scattering off of electrons do not currently exclude much of the parameter space of 
this model but it will be probed in further detail in the near future. The masses of the DP KK states in 4-D are essentially generated by our choice of BC, without the need of a dark Higgs field. 
When a scalar mediator field is present in the bulk the limits on invisible SM Higgs decays impose important constrains on the corresponding scalar couplings to the SM but only one of 
our BM is found to not satisfy this bound. We found that the thick brane mechanism considered here naturally reduces the Higgs portal couplings as well as the DP gauge couplings without the 
need of BC to force the vanishing of the Higgs portal in the vicinity of the SM fields (as described in Refs. \cite{Rizzo:2018ntg,Rizzo:2018joy}) and the requirement that the 
kinetic mixing is described by a small 
parameter.

Although it was worth to study the behavior of the Higgs portal coupling due to the addition of an extra scalar field in the bulk, it does not influence the constraints obtained for our DM candidate, thus the presence of this scalar is not mandatory in order for the model to be self-consistent.

Potential signatures of the present model include a combination of searches for UED-like particles and KK towers of the DP states. Missing energy searches at LHC, from the cascade 
decay of excited KK modes of the SM particles, constrain the compactification radius $L^{-1}$ as in the UED models. In addition to these searches, for heavier DP particles of the KK tower, which are at least 
twice as heavy as DM, the resulting cascade decay also gives a missing energy signature. For the lightest DP  (which is slightly heavier than DM) the main decay final states are likely to be charged  leptons or missing energy. Therefore, the present approach can lead to distinct signatures in ongoing and upcoming experiments, since it 
combines some common features of UED and DP models in an ED.

\acknowledgments

RGL is supported by CNPq  under the grant 208206/2017-5. The work of TGR was supported by the Department of Energy, Contract DE-AC02-76SF00515.


\bibliographystyle{JHEP}
\bibliography{references}


\end{document}